\documentclass[aip,12pt]{revtex4-1}
\usepackage{graphicx}
\usepackage{amsmath}
\usepackage{placeins}
\usepackage{multirow}
\begin{document}
\title{Disentangling relativistic spin torques  in a ferromagnet/semiconductor bilayer}
\author{T. D. Skinner}
\email[email: ]{timothy.skinner@ucl.ac.uk}
\altaffiliation{Present address: London Centre for Nanotechnology, University College London, WC1H 0AH, United Kingdom}
\affiliation{Cavendish Laboratory, University of Cambridge, CB3 0HE, United Kingdom}

\author{K. Olejn\'{i}k}
\affiliation{Institute of Physics, ASCR, v.v.i., Cukrovarnicka 10, 16253 Praha 6, Czech Republic}

\author{L. K. Cunningham}
\affiliation{Cavendish Laboratory, University of Cambridge, CB3 0HE, United Kingdom}

\author{H. Kurebayashi}
\altaffiliation{Present address: London Centre for Nanotechnology, University College London, WC1H 0AH, United Kingdom}
\affiliation{Cavendish Laboratory, University of Cambridge, CB3 0HE, United Kingdom}
\affiliation{PRESTO, Japan Science and Technology Agency, Kawaguchi 332-0012, Japan}

\author{R. P. Campion}
\affiliation{School of Physics and Astronomy, University of Nottingham, Nottingham NG7 2RD, United Kingdom}

\author{B. L. Gallagher}
\affiliation{School of Physics and Astronomy, University of Nottingham, Nottingham NG7 2RD, United Kingdom}

\author{T. Jungwirth}
\affiliation{Institute of Physics, ASCR, v.v.i., Cukrovarnicka 10, 16253 Praha 6, Czech Republic}
\affiliation{School of Physics and Astronomy, University of Nottingham, Nottingham NG7 2RD, United Kingdom}

\author{A. J. Ferguson}
\email[email: ]{ajf1006@cam.ac.uk}
\affiliation{Cavendish Laboratory, University of Cambridge, CB3 0HE, United Kingdom}

\date{\today}
\maketitle

\textbf{
Recently discovered relativistic spin torques induced by a lateral current at a ferromagnet/paramagnet interface are a candidate spintronic technology for a new generation of electrically-controlled magnetic memory devices\cite{Miron:2011_b,Liu:2012_a}. Phenomenologically, the torques have field-like and antidamping-like components with distinct symmetries. Microscopically, they are considered to have two possible origins. In one picture, a spin-current generated in the paramagnet via the relativistic spin Hall effect\cite{Jungwirth:2012_a} (SHE) is absorbed in the ferromagnet and induces the spin transfer torque\cite{Ralph:2007_a} (STT). In the other picture, a non-equilibrium spin-density  is generated  via the relativistic inverse spin galvanic effect\cite{Ivchenko:2008_a} (ISGE) and induces the spin-orbit torque\cite{Bernevig:2005_c,Manchon:2008_b,Chernyshov:2009_a} (SOT) in the ferromagnet. From the early observations in paramagnetic semiconductors, SHE and ISGE are known as companion phenomena that  can both allow for electrically aligning spins in the same structure\cite{Kato:2004_d,Kato:2004_b,Wunderlich:2004_a}. It is essential for our basic physical understanding of the spin torques at the ferromagnet/paramagnet interface to experimentally disentangle the SHE and ISGE contributions. To achieve this we prepared an epitaxial transition-metal-ferromagnet/semiconductor-paramagnet single-crystal structure and performed a room-temperature vector analysis of the relativistic spin torques  by means of the all-electrical ferromagnetic resonance (FMR) technique. By design, the field-like torque is governed by the ISGE-based mechanism in our structure while the antidamping-like torque is due to the  SHE-based mechanism.}

The splitting of the two microscopic mechanisms between the field-like and the antidamping-like torque components has not been previously achieved for several conceptual reasons. The original theoretical proposals\cite{Aronov:1989_a,Edelstein:1990_a,Mal'shukov:2002_a} and experimental observations\cite{Silov:2004_a,Kato:2004_b,Ganichev:2004_b,Wunderlich:2004_a} of the ISGE were made in paramagnets with no ferromagnetic component in the structure. The corresponding non-equilibrium spin-density, generated in the ISGE by inversion-asymmetry terms in the relativistic Hamiltonian, has naturally no dependence on magnetization. Hence, in the context of magnetic semiconductors\cite{Bernevig:2005_c,Chernyshov:2009_a,Endo:2010_a,Fang:2010_a} or ferromagnet/paramagnet structures\cite{Manchon:2008_b,Miron:2010_a,Pi:2010_a,Suzuki:2011_a,Miron:2011_b}, the ISGE may be expected to yield only the field-like component of the torque $\sim{\bf \hat{M}}\times\hat{\boldsymbol\zeta}$, where  the vector $\hat{\boldsymbol\zeta}$ is independent of the magnetization vector ${\bf \hat{M}}$ (see Fig.~1a). However, when carriers experience both  the spin-orbit coupling  and magnetic exchange coupling,  the inversion asymmetry can generate a non-equilibrium spin density component of extrinsic, scattering-related\cite{Pesin:2012_a,Wang:2012_b} or intrinsic, Berry-curvature\cite{Garate:2009_a,Kurebayashi:2013_a,Freimuth:2013_a} origin which is magnetization dependent and yields an antidamping-like torque $\sim{\bf \hat{M}}\times({\bf \hat{M}}\times\hat{\boldsymbol\zeta})$. Experiments in (Ga,Mn)As confirmed the presence of the ISGE-based mechanism\cite{Chernyshov:2009_a,Endo:2010_a,Fang:2010_a} and demonstrated that the field-like and the Berry-curvature antidamping-like SOT components can have comparable magnitudes\cite{Kurebayashi:2013_a}.

The STT is dominated by the antidamping-like component\cite{Ralph:2007_a} in weakly spin-orbit coupled ferromagnets with $\tau_{ex}\ll\tau_s$, where $\tau_{ex}$ is the precession time of the carrier spins in the exchange field of the ferromagnet and $\tau_s$ is the spin life-time in the ferromagnet. This, in principle, applies also to the case when the spin current is injected to the ferromagnet from a paramagnet via the SHE (see Fig. 1b). However, at finite $\tau_s$, the STT also acquires a field-like component\cite{Ralph:2007_a}. Experiments in W/Hf/CoFeB structures confirmed the presence of the SHE-based mechanism in the observed torques and showed that the SHE-STT can have both antidamping-like and field-like components of comparable magnitudes\cite{Pai:2014_a}.

In the commonly studied polycrystalline  transition-metal-ferromagnet/heavy-metal-paramagnet samples, the dependence of the torques on the angle of the driving in-plane current also does not provide the direct means to disentangle the two microscopic origins. The lowest order inversion-asymmetry spin-orbit terms in the Hamiltonian have the Rashba form for which  the vector $\hat{\boldsymbol\zeta}$ is in the plane parallel to the interface and perpendicular to the current, independent of the current direction. The same applies to the spin-polarization of the SHE spin-current propagating from the paramagnet to the ferromagnet. The ${\bf \hat{M}}$ and $\hat{\boldsymbol\zeta}$ functional form of the field-like and antidamping-like SHE-STTs is the same as of the corresponding SOT components. In the observed lowest order torque terms in Pt/Co and Ta/CoFeB structures\cite{Garello:2013_a} the ISGE-based and the SHE-based mechanism remained, therefore, indistinguishable. The simultaneous observation of  higher order torque terms in these samples pointed to SOTs due to structural inversion-asymmetry terms beyond the basic Rashba model. From the Ta thickness dependence measurements in the Ta/CoFeB structure it was concluded that  in these samples both the ISGE-based and the SHE-based mechanisms contributed to both the field-like and the antidamping-like torques\cite{Kim:2013_a} . 

The SHE and ISGE were originally discovered in III-V semiconductors\cite{Kato:2004_d,Wunderlich:2004_a,Silov:2004_a,Kato:2004_b,Ganichev:2004_b} but for maximizing the relativistic spin torques in common transition-metal ferromagnets it turned out more suitable to interface them with the highly conductive heavy-metal paramagnets like Pt, Ta, or W\cite{Miron:2011_b,Liu:2012_a,Kim:2013_a,Garello:2013_a,Pai:2014_a}. Rather than enhancing the magnitude of the torques, the aim of our study is to clearly separate the two microscopic origins for which returning to the III-V semiconductor paramagnets is instrumental. In experiments presented below we succeed to  disentangle the torques into the ISGE driven field-like component and the SHE driven antidamping-like component using a  2~nm thick single crystal Fe grown epitaxially without breaking vacuum on top of a 20~nm thick p-doped GaAs epilayer (see Methods). 

Among the inversion-asymmetry terms characteristic of our system, the Rashba term due to the structural asymmetry of the bilayer plays a minor role. Instead, the leading role is played by the broken inversion symmetry in the crystal structure of the  semiconductor paramagnet. Independent of the interface, holes in the strained zinc-blende  lattice of the III-V semiconductor experience a linear-in-wavevector Dresselhaus spin-orbit coupling. The ISGE of the corresponding symmetry generated in the semiconductor induces the SOT in the adjacent Fe film with $\hat{\boldsymbol\zeta}$ perpendicular to the current for current along [110] or [1$\bar{1}$0] crystal axes of the semiconductor, while $\hat{\boldsymbol\zeta}$ is parallel to the current for the [100] or [010] current directions. Moreover, the ISGE with this  characteristic Dresselhaus symmetry generated in the semiconductor paramagnet induces only the field-like torque in the adjacent Fe while the antidamping-like SOT component with this symmetry is absent by design.  On the other hand, the SHE spin-current can be readily generated inside the paramagnetic p-doped GaAs layer and absorbed in the weakly spin-orbit coupled Fe in the form of the antidamping-like STT. 

To obtain measurable torques in a metal-ferromagnet/semiconductor-paramagnet structure requires at least partially matched conductances of the semiconductor and metal layers  which we achieved by doping the GaAs host with  $\sim 3$\% of  substitutional Mn$_{\rm Ga}$ acceptors. Mn allows us to achieve this exceptionally high charge-doping. Simultaneously, at 3\% Mn, hole bands within the (Ga,Mn)As semiconductor do not experience a sizeable enough magnetic exchange splitting at room temperature that would generate an observable Dresselhaus-symmetry antidamping-like SOT in our structure.

We have measured the relativistic spin torques  using the electrically induced and detected FMR technique\cite{Fang:2010_a,Fang:2012_a} (see Fig.~2a). In this method, a microwave current flowing in the device induces FMR when the externally applied magnetic field matches the resonant condition. The resonance of the Fe magnetisation can be detected in the dc voltage induced across the bar, $V_{\mathrm{dc}}$. This is due to the homodyne mixing of the microwave current with the oscillating component of magnetoresistance caused by the magnetisation precession. In these measurements we increase the microwave current coupled into the sample, with a typical resistance of 8 k$\Omega$, by using an impedance matching network\cite{Fang:2012_a}.

Current-induced torques were measured in $10\times200$~ $\mu$m patterned bars. For a series of external magnetic field directions in the plane of the sample, FMR sweeps were recorded using a microwave frequency of close to 16 GHz. From the magnetisation angle-dependence of the resonance field, we obtained the magnetisation amplitude value of $\mu_0 M_{\rm s}=1.85\pm 0.03$ T which is close to the literature value of 1.7 T for Fe\cite{Danan:1968_a}, and we found an in-plane uniaxial anisotropy of $\mu_0$H$_{\mathrm{U}}=0.101\pm0.001$ T which is typical for thin films of Fe grown on GaAs\cite{Brockmann:1999_a}. By solving the Landau-Lifshitz-Gilbert equation for a small current induced excitation field, $(h_{\mathrm{x}},h_{\mathrm{y}},h_{\mathrm{z}})e^{j\omega t}$, $V_{\mathrm{dc}}$ is found to be comprised of symmetric and antisymmetric Lorentzian functions with coefficients $V_{\mathrm{sym}}$ and $V_{\mathrm{asy}}$ respectively. Here $h_x$ is the excitation field component parallel to the current and $h_y$ is the component perpendicular to the current. The coefficients depend on the angle, $\theta$, of the magnetisation vector relative to the current and are given by
\begin{align}\label{eq:Vdc}
V_{\mathrm{sym}}&=V_{\mathrm{mix}}A_{\mathrm{yz}}\sin 2\theta h_{\mathrm{z}},\\\nonumber
V_{\mathrm{asy}}&=V_{\mathrm{mix}}A_{\mathrm{yy}}\sin 2\theta(h_{\mathrm{y}}\cos\theta - h_{\mathrm{x}}\sin\theta).
\end{align}
Here, $V_{\mathrm{mix}}$ is the sensitivity of the mixing detection and is given by $V_{\mathrm{mix}}=-\tfrac{1}{2}I_{0}\Delta R$, where $I_{0}\left(e^{j\omega t},0,0\right)$ is the microwave current in the device and $\Delta R$ is the coefficient of the anisotropic magnetoresistance of the sample. $A_{\mathrm{yy}}$ and $A_{\mathrm{yz}}$ are the diagonal and off-diagonal components of the ac magnetic susceptibility, which depend on the magnetic anisotropies and Gilbert damping of the sample. In our devices, $\Delta R$ is typically 17~$\Omega$ which, assuming  Fe carries the majority of the current in the bilayer, is consistent in sign and magnitude with literature values of 0.2\% anisotropic magnetoresistance in Fe.\cite{McGuire:1975_a} We estimate the proportion of total bilayer current in the Fe layer to be 79\% by resistance measurements of Hall bars before and after removing the Fe and capping Al (see supplementary information).

FMR measurements were made using devices patterned in four crystal directions. The microwave power for all devices, incident on the impedance matching network, was 24 dBm. For each angle, the resonances were fitted by symmetric and antisymmetric Lorentzian functions. A typical curve is shown in Fig. \ref{fig2}b. The in-plane uniaxial magnetic anisotropy of Fe implies that, in general, the magnetisation does not lie along the external field. The actual magnetisation angles and uniaxial anisotropy are self-consistently calculated from the dependence of the resonant field on the external magnetic field angle\cite{Ando:2008_c}. This also allows the susceptibilities, $A_{\mathrm{yz}}$ and $A_{\mathrm{yy}}$, to be calculated. The magnetisation angle dependence of $V_{\mathrm{sym}}/A_{\mathrm{yz}}$ and $V_{\mathrm{asy}}/A_{\mathrm{yy}}$ is plotted in Fig. \ref{fig2}c for a bar patterned in the [010] crystal direction. 

Our analysis of the current-induced torques using equation (\ref{eq:Vdc}) is not necessarily valid if the torques do not act in phase with the microwave current. In comparison to electrically detected FMR measurements where the microwave current is capacitively or inductively coupled into the sample\cite{Harder:2011_a}, we do not expect a phase-shift between the microwave current and induced fields as the current is conducted ohmically. Nevertheless, we might worry that some part of our microwave resonator circuit leads to a phase shift. To test this, we repeated our measurements with a [100] device over a frequency range (11.8 to 14.4 GHz) using a microstrip resonator with a fundamental frequency close to 13 GHz (Fig. \ref{fig3}a). If there were some frequency dependent phase shift, we would expect the lineshape to oscillate between an antisymmetric and symmetric Lorentzian over this frequency range. However, the ratio of $V_{\mathrm{sym}}$ to $V_{\mathrm{asy}}$ remains constant in this frequency range to within experimental error (Fig. \ref{fig3}b), confirming that our analysis is correct.

As shown in Fig.~\ref{fig4}a, the in-plane current-induced field depends strongly on the crystal direction of the current and can be well fitted by the Dresselhaus-symmetry ISGE field, ${\bf h}^{\mathrm{ISGE}}\sim\left(\cos 2\phi_{[100]},-\sin 2\phi_{[100]},0\right)$, where $\phi_{[100]}$ is the angle between the current and the [100] crystal direction. This is the expected symmetry of the current-induced non-equilibrium spin-polarisation of carriers in the semiconductor due to the inversion-asymmetric crystal structure of the strained zinc-blende lattice of (Ga,Mn)As. The interface exchange coupling of these polarized carriers with the adjacent Fe moments induces the field-like SOT in Fe with the Dresselhaus symmetry. We note that other torque terms with the symmetry common to the Rashba ISGE, the field-like component of the SHE-STT, or the torque due to an Oersted field have only a minor contribution to the total measured field-like torque.

To highlight that carriers in the semiconductor layer are responsible for the Dresselhaus-symmetry ISGE field, we compare Fig.~\ref{fig4}a with previous measurements in which the in-plane current induced fields were measured in a bare (Ga,Mn)As epilayer\cite{Kurebayashi:2013_a} without the Fe film. To observe the corresponding SOT in this sample, a larger concentration of magnetic Mn-moments was used, and the measurements were performed at low temperatures where the Mn moments are ferromagnetic in equilibrium. Instead of the interfacial exchange coupling to Fe, the current induced non-equilibrium spin-polarisation of carriers in the semiconductor due to the Dresselhaus-symmetry ISGE is exchange-coupled directly to the ferromagnetic moments on which it exerts the field-like SOT. In both the Fe/(Ga,Mn)As and the (Ga,Mn)As samples the same crystal-symmetry field-like SOT is observed which confirms their common Dresselhaus ISGE origin.

In contrast to the in-plane field, the out-of-plane current induced field is independent of the crystal direction of the current but depends on the magnetisation angle. It is dominated by a term ${\bf h}^{\mathrm{SHE}}\sim{\bf \hat{M}}\times{\bf \hat{y}}$ (${\bf \hat{y}}$ is the direction perpendicular to the current) which generates the antidamping-like torque. As shown in Fig.~\ref{fig4}b, the amplitudes of the field-like and antidamping-like torques are comparable in our Fe/(Ga,Mn)As structure. The underlying microscopic mechanism of the antidamping-component can only be of the SHE-STT origin. 

In previous measurements in the bare (Ga,Mn)As epilayer,\cite{Kurebayashi:2013_a} the antidamping-like SOT was dominated by the counterpart microscopic mechanism to the Dresselhaus ISGE. This Dresselhaus-symmetry antidamping-like SOT is clearly missing in our measured data. It is suppressed in our Fe/(Ga,Mn)As structure by design because carriers in the semiconductor are not sufficiently magnetized at equilibrium due to the low Mn moment density and high temperature of the experiment. 

A Rashba-symmetry antidamping-like SOT due to the carriers in the Fe experiencing the inversion-asymmetry of the interface could in principle also explain our measured data. This antidamping SOT would have the same symmetry as the antidamping SHE-STT. This possibility is, however, ruled out by our control experiment in which we perform electrically detected FMR in a similar MBE-grown Fe (1 nm)/insulating GaAs structure at room temperature. In this case, we do not observe the anti-damping torque in the rectification effect, despite the sample possessing a similar magnetoresistance ratio ($\sim$0.2\%) to our Fe/(Ga,Mn)As. This is consistent with the carriers being removed from the semiconductor which eliminates the SHE source of the spin-current. We note that also consistently with the absence of carriers in the semiconductor in the Fe/insulating-GaAs structure, we do not observe the Dresselhaus-symmetry field-like SOT in this control sample. 

To calibrate the microwave current in the sample we used a bolometric technique  (see supplementary information).  Using this calibration, we estimate amplitudes of $|\mu_0h^{\mathrm{ISGE}}/J_{\mathrm{GaAs}}|=16\pm10$ $\mu$T/10$^6$Acm$^{-2}$ and $|\mu_0h^{\mathrm{SHE}}/J_{\mathrm{GaAs}}|=20\pm9$ $\mu$T/10$^6$Acm$^{-2}$. The error is found from the statistical variation from all of the devices measured. To verify the bolometric calibration, we also perform an additional check with a single device by measuring the change in Q-factor of the microstrip resonator loaded with and without a sample (see supplementary information). This calibration yields values of $|\mu_0h^{\mathrm{ISGE}}/J_{\mathrm{GaAs}}|=37$ $\mu$T/10$^6$Acm$^{-2}$ and $|\mu_0h^{\mathrm{SHE}}/J_{\mathrm{GaAs}}|=47$ $\mu$T/10$^6$Acm$^{-2}$, close to the values of the bolometric technique.

From the measured $h^{\mathrm{SHE}}$ in our Fe/(Ga,Mn)As structure we can infer the room-temperature spin Hall angle, $\theta_{\mathrm{SH}}$, in the paramagnetic (Ga,Mn)As using the expression based on the antidamping-like STT\cite{Liu:2012_a}, 
\begin{equation}
\theta_{\mathrm{SH}}=\dfrac{2e}{\hbar}\mu_0 M_{\mathrm{s}}d_{\mathrm{Fe}}\dfrac{h^{\mathrm{SHE}}}{J_{\mathrm{GaAs}}}.
\label{SHEangle}
\end{equation}
Here it is assumed that the thickness of the semiconductor is much larger than its spin diffusion length (5.6~nm in p-GaAs\cite{Chen:2013_a}) and $d_{\mathrm{Fe}}$ is the thickness of the Fe layer. Eq.~(\ref{SHEangle}) yields values of $\theta_{\mathrm{SH}}=1.7\pm 0.9$\% (bolometric calibration) and 4\% (Q-factor calibration), similar to spin Hall angles previously reported for carriers with p-orbital character in GaAs\cite{Chen:2013_a, Okamoto:2014_a}. To check these spin-Hall values we compare our torque efficiencies in terms of field per total current density with those of transition-metal/ferromagnet bilayers. For instance, Garello \textit{et al.} measured an antidamping-like torque of $|\mu_0h/J|=690$ $\mu$T/10$^6$Acm$^{-2}$ in layers with 0.6 nm Co and 3 nm Pt and an equivalent spin-Hall angle of 16\%.\cite{Garello:2013_a} Although our spin-Hall angle is only 4-8 times smaller, the field per total current density is 85-170 times smaller because the total magnetic moment of the Fe is $\sim4$ times higher than of the Co and $\sim80$\% of the total current is shunted through the Fe. 

To conclude, we have experimentally disentangled the two archetype microscopic mechanisms that can drive relativistic current-induced torques in ferromagnet/paramagnet structures. In our epitaxial Fe/(Ga,Mn)As bilayer we simultaneously observed ISGE-based and SHE-based torques of comparable amplitudes. Designed magnetization-angle and current-angle symmetries of  our single-crystal structure allowed us to split the two microscopic origins between the field-like and the antidamping-like torque components. Experimentally establishing the microscopic physics of the relativistic spin torques  should stimulate both the fundamental and applied research of these intriguing and practical spintronic phenomena.

\medskip
\textbf{Methods}

The semiconductor (Ga,Mn)As layer of thickness 20~nm was deposited on a GaAs(001) substrate at a temperature of 260$^\circ$C. The substrate temperature was then reduced to 0$^\circ$C, before depositing a 2~nm Fe layer, plus a 2~nm Al capping layer. In-situ reflection high energy electron diffraction and ex-situ x-ray reflectivity and diffraction measurements confirmed that the layers are single-crystalline with sub-nm interface roughness. 

To improve the sensitivity of the FMR measurement, the sample is embedded in a microstrip resonator circuit\cite{Fang:2012_a} which acts to impedance match the $\sim$ 8 k$\Omega$ sample to the external 50 $\Omega$ transmission line at the fundamental frequency (in this case $\sim$ 8 GHz) or harmonic frequencies of the resonator. To allow measurement of $V_{\mathrm{dc}}$, the resonator contains an on-board bias-T. In this experiment, FMR measurements are made at the the 2$^{\text{nd}}$ harmonic frequency of the resonator ($\sim$ 16 GHz).

\medskip
\textbf{Acknowledgements}

The authors acknowledge support from EU European Research Council (ERC) advanced grant no. 268066, from the Ministry of Education of the Czech Republic grant no. LM2011026, from the Grant Agency of the Czech Republic grant no. 14-37427G and the Academy of Sciences of the Czech Republic Praemium Academiae. A.J.F. acknowledges support from a Hitachi research fellowship. H.K. acknowledges financial support from the Japan Science and Technology Agency (JST).

\medskip
\textbf{Author Contributions}

K.O., R.P.C., B.L.G. and H.K. grew the materials. K.O. and H.K. prepared the samples. T.D.S., L.K.C. and H.K. performed the experiments and T.D.S. analysed the data. T.D.S., T.J. and A.J.F. prepared the manuscript. T.D.S. and A.J.F. planned the project.


\FloatBarrier
\begin{figure}
\centering
\includegraphics[]{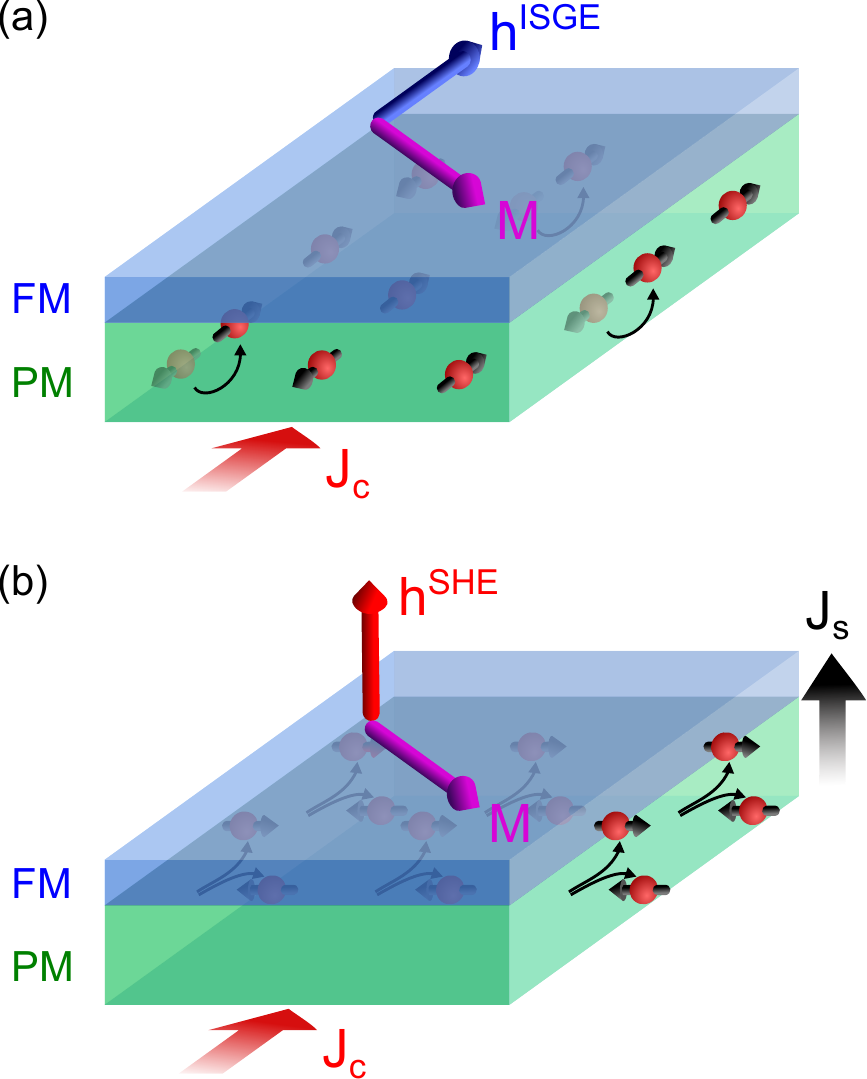}
\caption{(a) In the GaAs layer, as charge carriers are accelerated by an applied electric field in the inversion-asymmetric crystal potential, scattering events lead to a non-equilibrium spin-polarisation. The spin-polarisation depends on the direction of the current with respect to crystal direction. Through exchange coupling at the interface, the magnetisation experiences an effective field, $h^{\mathrm{ISGE}}$, parallel to the spin-polarisation.
(b) In the GaAs layer, when a longitudinal charge current is applied, a transverse spin-current is induced by the SHE, which flows into the Fe layer. The spin-current exerts a torque that depends on the magnetisation angle. For an in-plane magnetisation this is described by an out-of-plane effective field, $h^{\mathrm{SHE}}$.}
\label{fig1}
\end{figure}

\begin{figure}
\centering
\includegraphics[]{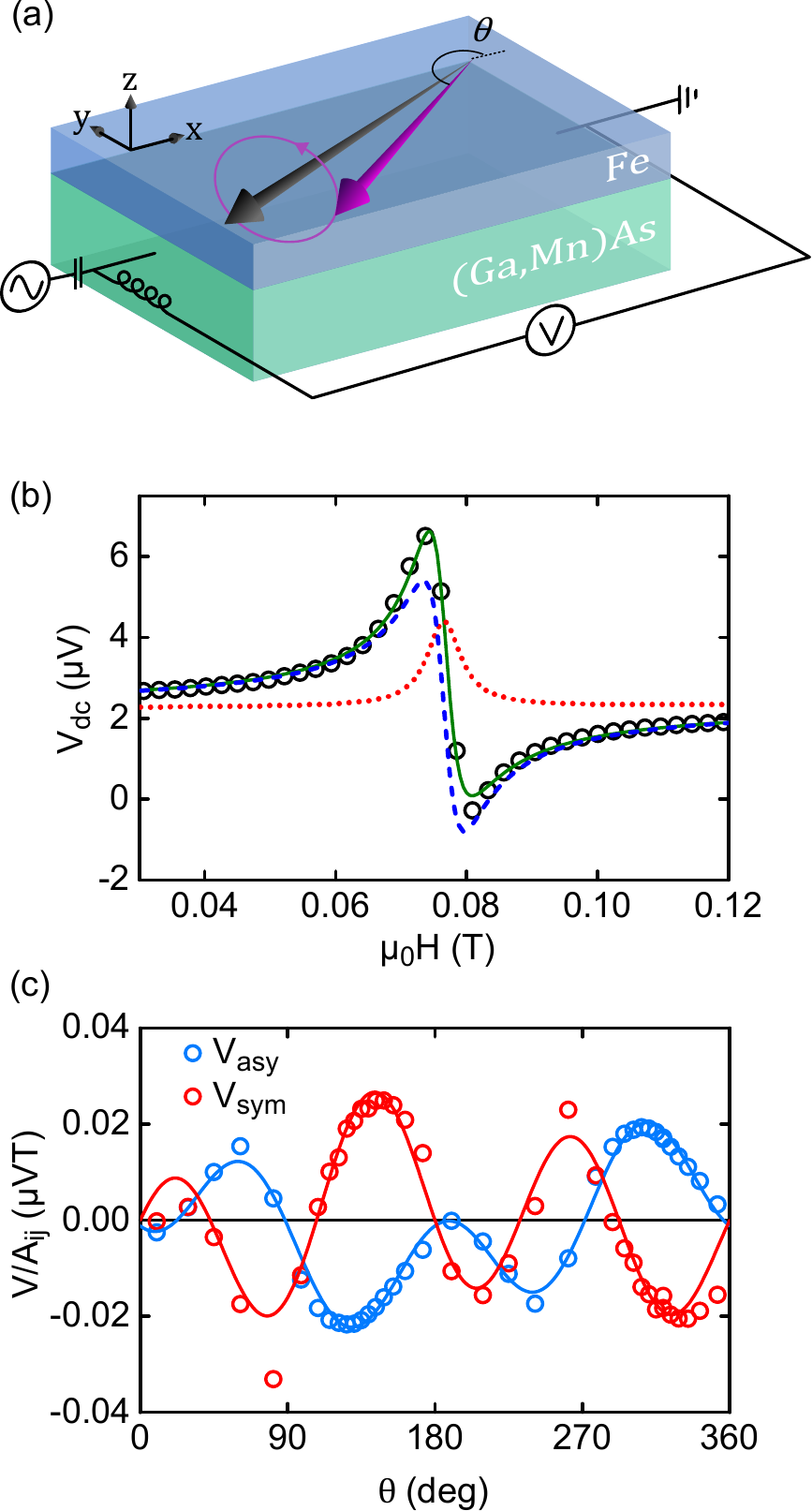}
\caption{(a) Schematic of the measurement showing magnetisation precession. A microwave current passes through the capacitor of the bias-T and into the Fe/GaAs bilayer sample where the current-induced torques causes the magnetisation of the Fe to precess around the external field. The precession is detected in the dc voltage measured across the device by a magnetoresistance mixing effect. (b) A typical SO-FMR curve detected in $V_{\mathrm{dc}}$ as a function of external field, induced by a 16.245 GHz microwave current.  $V_{\mathrm{dc}}$ is fitted by a combination of symmetric (red dotted line) and antisymmetric (blue dashed line) Lorentzians. (c) The in-plane magnetisation angle dependence of the fitted Lorentzian amplitudes for a device with current in the [010] crystal direction. The full expression for the angle-dependence of the fitted data is given in the supplementary information.}\label{fig2}
\end{figure}

\begin{figure}
\centering
\includegraphics[]{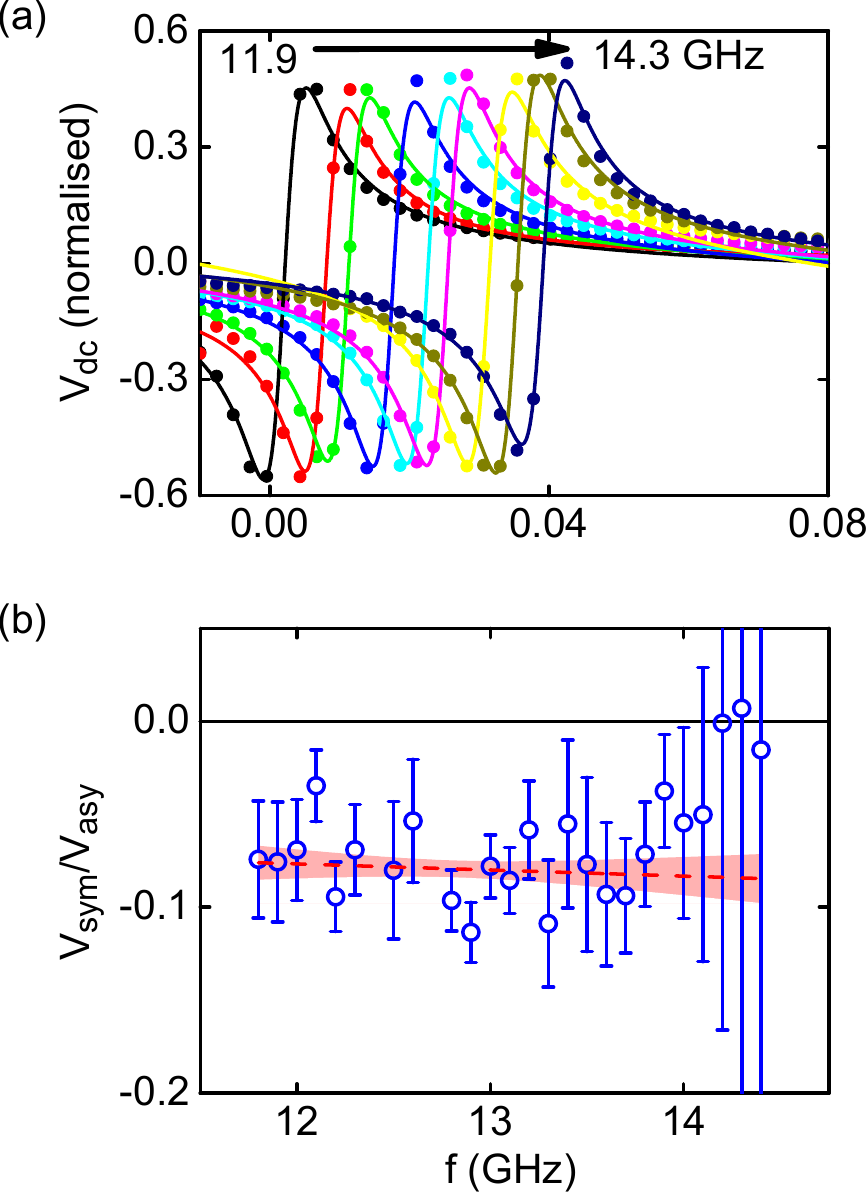}
\caption{(a) Normalised, fitted, resonant peaks for a series of microwave frequencies detected in $V_{\mathrm{dc}}$. The line-shape is dominated by an antisymmetric Lorentzian for every frequency. (b) The ratio of $V_{\mathrm{sym}}$ to $V_{\mathrm{asy}}$ (blue circles) is constant to within the the standard error of a linear fit (red dashed line) for the measured frequency range. The error in the ratio is bigger away from the impedance-matched frequency (13 GHz) of the microstrip resonator used as the detected peak in $V_{\mathrm{dc}}$ is smaller.}\label{fig3}
\end{figure}

\begin{figure}
\centering
\includegraphics[]{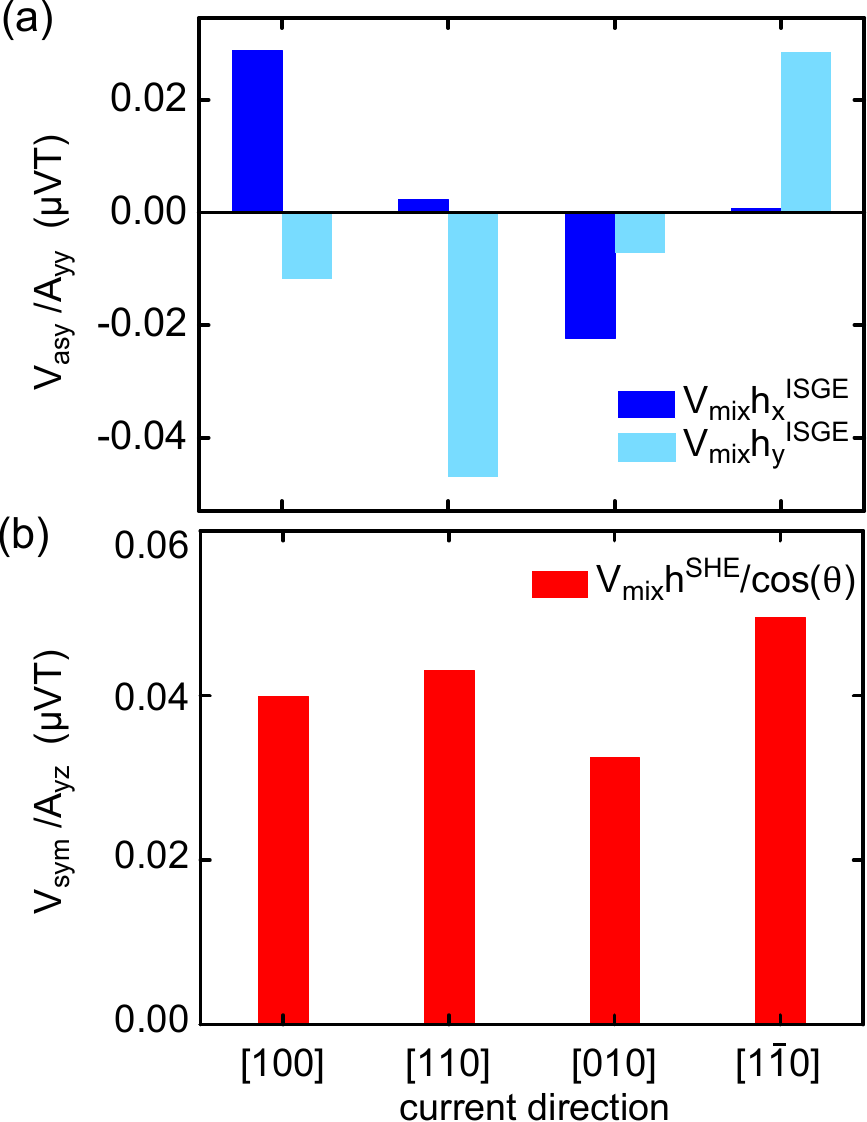}
\caption{(a) The fitted in-plane field coefficients for a set of devices in different crystal directions. (b) The fitted out of plane field coefficient (representing the antidamping torque) for the same devices.}\label{fig4}
\end{figure}

\FloatBarrier
\pagebreak
\begin{center}
\textbf{\large Supplementary Information}
\end{center}
\setcounter{equation}{0}
\setcounter{figure}{0}
\setcounter{table}{0}
\setcounter{page}{1}
\makeatletter
\renewcommand{\theequation}{S\arabic{equation}}
\renewcommand{\thefigure}{S\arabic{figure}}
\renewcommand{\thetable}{S\arabic{table}}
\renewcommand{\thesection}{S\arabic{section}}
\renewcommand{\bibnumfmt}[1]{[S#1]}
\renewcommand{\citenumfont}[1]{S#1}

\section{Impedance matching network}
We use a gap-coupled microstrip resonator, similar to one previously reported\cite{Fang:2012_a}, to impedance match the $\sim$ 8 k$\Omega$ sample to the 50 $\Omega$ transmission line. In contrast to the network previously reported, the sample is attached to the opposite end of the resonator with a wirebond. At the resonant frequency of the resonator, the impedance of the network becomes real and is given by
\begin{equation}
Z \approx \dfrac{1}{R \omega^2 C_{\mathrm{k}}^2}.
\end{equation}
$C_{\mathrm{k}}$ is the capacitance of the interdigitated gap capacitor and R is the resistance of the sample. By choosing a suitable value of $C_{\mathrm{k}}$, at the resonant frequency $Z \approx 50 \Omega$ and most of the microwave power is transmitted from the transmission line to the matched network. For the FMR measurements, we use the resonator at twice its fundamental frequency, $\sim$ 16 GHz, where the power is also matched to the sample. To measure $V_{\mathrm{dc}}$, a wirebond is attached at $\frac{1}{4}$ of the length of the resonator. This causes little perturbation of the resonator mode, as at this point a node of electric field exists. 

\section{Q factor calibration of microwave current}
Close to the resonance, the microstrip resonator can be approximated as a series resonator. The quality factor, $Q$, of the resonator can then be described by
\begin{equation}
\dfrac{1}{Q} = \dfrac{1}{Q_{\mathrm{sample}}} + \dfrac{1}{Q_{\mathrm{board}}},
\end{equation}
where $Q_{\mathrm{sample}}$ represents $Q$ due to power dissipated in the sample and $Q_{\mathrm{board}}$ represents the $Q$ due to power dissipated through losses in the PCB. When connected to an external network, the effect of dissipation externally can be described by a total $Q$ factor
\begin{equation}
Q_{\mathrm{total}}=\dfrac{Q}{\left(1+g\right)},
\end{equation}
where g is a coupling constant that describes the impedance matching of the resonator, given by $g=Z_0/Z$ and $Z_0 = 50$ $\Omega$ is the the characteristic impedance of the transmission line.

Both Q and g can be found from measuring the reflected power from the resonator network around the resonant frequency. At resonance, the reflection coefficient is given by 
\begin{equation}
|\Gamma|^2=\left(\dfrac{1-g}{1+g}\right)^2.
\end{equation}
We can also determine the dissipation of the resonant circuit through the width of the absorption peak, with $Q_{\mathrm{total}}$ given by 
\begin{equation}
Q_{\mathrm{total}} = \dfrac{\omega}{\Delta \omega_{\mathrm{FWHM}}}
\end{equation}
where $\Delta \omega_{\mathrm{FWHM}}$ is the full width at half maximum of the absorption peak.

\begin{figure}
\includegraphics[width=0.5\textwidth]{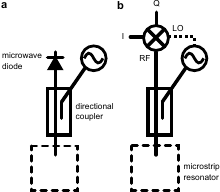}
\caption{Microwave set-ups for measuring (a) the reflected power and (b) the reflected phase from the microstrip resonator network.}\label{suppl2}
\end{figure}

To perform a calibration of the microwave current in a sample, we use a directional coupler and calibrated microwave diode (Fig. \ref{suppl2}a) to measure the reflected signal with and without the resonator loaded by the sample. If we replace the diode with a microwave mixer (Fig. \ref{suppl2}b), we can measure the phase change of the reflected signal around the resonance of the microstrip resonator. We use the result that at resonance, the gradient of phase is given by
\begin{equation}
\left.\dfrac{\partial\phi}{\partial\omega}\right|_{\omega_{0}} = \dfrac{4Q}{\omega_0}\dfrac{g}{(1-g^2)},
\end{equation}
which allows us to determine the value of $Q$ when the resonator is not loaded by the sample and the resonance is hard to observe in reflected power. We calibrate a microstrip resonator at its fundamental frequency, as this has previously been studied. To achieve a high enough frequency for our FMR measurements, we reduce the length of the resonator so that the fundamental frequency is around 13 GHz.

\begin{figure}
\includegraphics[width=0.45\textwidth]{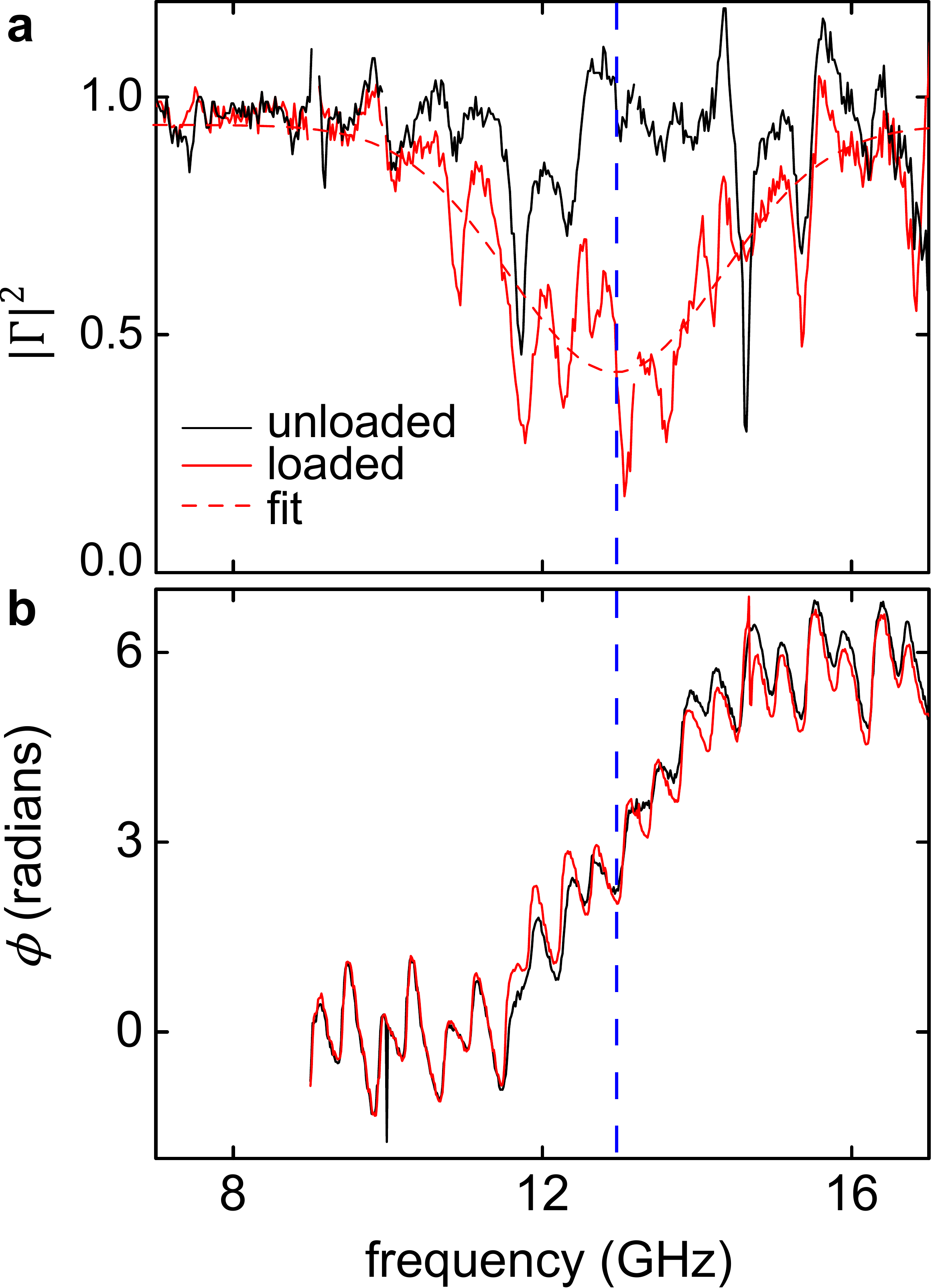}
\caption{(a) magnitude and (b) phase of the reflected power showing a resonance at around 13 GHz (blue dashed line) for the loaded and unloaded resonator. The full $2\pi$ phase change shows in both cases the resonator is overcoupled ($g>1$). The visible absorption peak in reflected power when the resonator is loaded with the sample shows that the microwave power is better matched at resonance.}\label{suppl1}
\end{figure}

The measured magnitude and phase of the reflected signal is shown in Fig. \ref{suppl1}a and b respectively. From the peak in absorption we estimate that for the loaded resonator, $g=4.3$. Although we cannot make out a clear peak from the noise for the unloaded case, we can put a lower limit on $g>20$. From the gradient of the phase and the calculated values for $g$, we find that $Q_{\mathrm{sample}}<<Q_{\mathrm{board}}$ and so assume all of the power transmitted to the microstrip resonator is dissipated in the sample. On resonance, using the reflection measurements we then estimate, for a 10 dBm source power, that the microwave current in the device is 0.74 mA. 

We perform FMR measurements at 10 dBm in a [100] bar as discussed in the main text and, using the value of the calibrated microwave current, find values of $|\mu_0h^{\mathrm{ISGE}}/J_{\mathrm{GaAs}}|=37$ $\mu$T/10$^6$Acm$^{-2}$ and $|\mu_0h^{\mathrm{SHE}}/J_{\mathrm{GaAs}}|=47$ $\mu$T/10$^6$Acm$^{-2}$.

\section{Joule heating calibration of microwave current}
\begin{figure}
\includegraphics[width=0.6\textwidth]{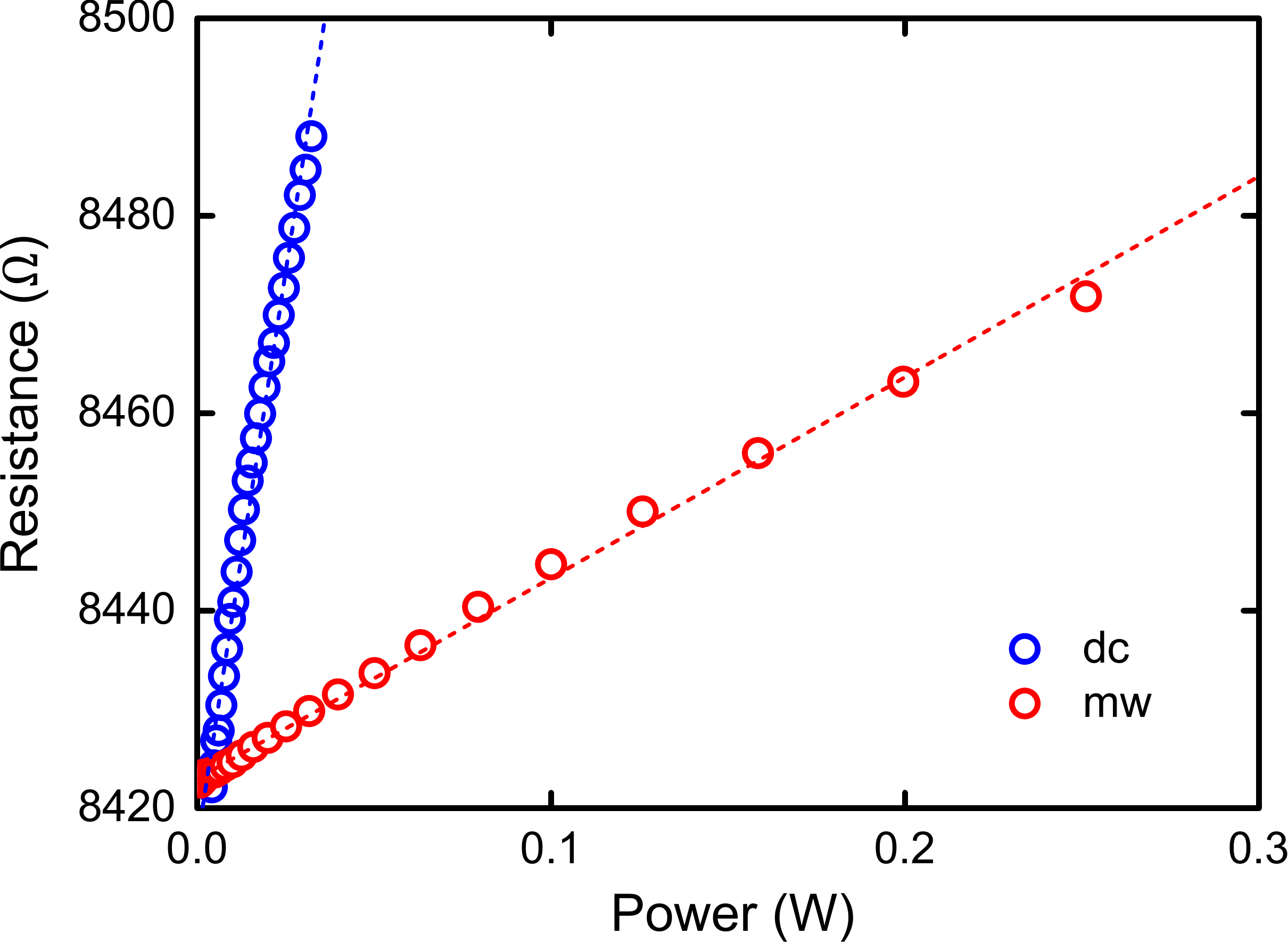}
\caption{Joule heating calibration for a [010] bar with 20 nm (Ga,Mn)As, showing the dependence of device resistance on dc and source microwave power. The ratio of the gradients (dashed lines) allows the fraction of transmitted microwave power to be calculated.}\label{suppl3}
\end{figure}

The microwave current can also be calibrated by a simple Joule heating method as the resistance of the semiconductor is sensitive to temperature. This calibration was performed for each device measured. To determine the microwave current in the device for a given source power, the change in resistance due to the heating of the microwave current is compared to the change due to heating of dc current. 

The dc current is applied from the bias-T of the resonator, through the sample, to ground. An applied dc voltage is held for 10 seconds at increasing values before the current is recorded. The resistance is then calculated as a function of dc current. Microwave power is then applied at increasing increments and held for 10 seconds at each value as before, before the resistance is measured from a small dc bias applied concurrently. We then compare the gradients of resistance with microwave and dc power (Fig. \ref{suppl3}). From the ratio of these gradients we estimate the proportion of microwave power dissipated in the sample.
\FloatBarrier

\section{FMR fitting coefficients}
\FloatBarrier
Our analysis in the main text only considers the dominant field components found from fitting the magnetisation angle dependence of $V_{\mathrm{dc}}$. The actual expression for the fitted angle dependence for all the samples is given by
\begin{align}\label{eq:fitting}
\dfrac{V_{\mathrm{sym}}}{A_{\mathrm{yz}}}&=C_{\sin\theta}\sin\theta + V_{\mathrm{mix}}\left(h_{\mathrm{z}}^0 + h_{\mathrm{z}}^{\cos\theta}\cos\theta + h_{\mathrm{z}}^{\sin\theta}\sin\theta\right)\sin 2\theta,\\\nonumber
\dfrac{V_{\mathrm{asy}}}{A_{\mathrm{yy}}}&=C_0 + V_{\mathrm{mix}}\left(h_{\mathrm{y}}\cos\theta - h_{\mathrm{x}}\sin\theta\right)\sin 2\theta + C_{\sin 2\theta}\sin 2\theta.
\end{align}
$C_{\sin\theta}$, $C_0$ and $C_{\sin 2\theta}$ are small additional coefficients which are empirically needed to fit the data. Furthermore, $h_{\mathrm{z}}^{\cos\theta}$ and $h_{\mathrm{z}}^{\sin\theta}$ are magnetisation angle dependent coefficients of the $h_{\mathrm{z}}$ effective field. The dominant effective fields in each of the samples are the in-plane fields, $h_{\mathrm{x}}$ and $h_{\mathrm{y}}$, originating from the ISGE and the magnetisation angle dependent out-of-plane field, $h_{\mathrm{z}}^{\cos\theta}$, which corresponds to the anti-damping SHE-STT. The analysis in the main text only considers these terms. Compared to the notation in the main text, $|{\bf h}^{\mathrm{SHE}}|\equiv h_{\mathrm{z}}^{\cos\theta}\cos{\theta}$ and $h_{\mathrm{x}}$ and $h_{\mathrm{y}}$ are the vector components of ${\bf h}^{\mathrm{ISGE}}$.

For completeness we present here  a table of all the components as well as the AMR coefficient and calibrated microwave current used to determine $V_{\mathrm{mix}}$. Table \ref{tab:20nm} shows the values for all of the devices.
\begin{table}[]
\centering
\begin{tabular}{cccccc}
\hline
&&\multicolumn{4}{c}{Direction}\\
\cline{3-6}
&& [100] & [110] & [010] & [1$\bar{1}$0] \\ 
\hline 
\multirow{4}{*}{$V_{\mathrm{asy}}/A_{\mathrm{yy}}$ (nVT)}&\multicolumn{1}{c|}{$\mu_0h_{\mathrm{x}}^{\mathrm{ISGE}}V_{\mathrm{mix}}$}&28.8 & 2.4 & -22.3 & 0.8\\ 
&\multicolumn{1}{c|}{$\mu_0h_{\mathrm{y}}^{\mathrm{ISGE}}V_{\mathrm{mix}}$}&-11.8 & -47.0 & -7.1 & 28.5\\
&\multicolumn{1}{c|}{$C_0$}&-2.6&-4.7&-1.2&-1.1\\
&\multicolumn{1}{c|}{$C_{\sin 2\theta}$}&5.0&-15.8&0.2&0.8\\
\hline
\multirow{4}{*}{$V_{\mathrm{sym}}/A_{\mathrm{yz}}$ (nVT)}&\multicolumn{1}{c|}{$\mu_0h^{\mathrm{SHE}}V_{\mathrm{mix}}/\cos{\theta}$}&39.9 & 43.1 & 32.5 & 49.6\\
&\multicolumn{1}{c|}{$\mu_0h_{\mathrm{z}}^0V_{\mathrm{mix}}$}&16.4 &-4.3 &-3.7 &2.2\\
&\multicolumn{1}{c|}{$\mu_0h_{\mathrm{z}}^{\sin\theta}V_{\mathrm{mix}}$}&4.4 &12.9 &13.6 &-2.9\\
&\multicolumn{1}{c|}{$C_{\sin \theta}$}&-19.8 &-4.5 &-16.1 &-16.1\\
\hline
&\multicolumn{1}{c|}{$\Delta R$ ($\Omega$)}& 17.7 & 15.4 & 17.5 & 16.5\\
&\multicolumn{1}{c|}{$I_0$ (mA)}& 3.5 & 2.2 & 1.8 & 1.8\\
\hline
\end{tabular}
\caption{Measurement parameters and all fitted detection coefficients of the Fe(2~nm)/Ga$_{0.097}$Mn$_{0.03}$As (20~nm) devices.}
\label{tab:20nm}
\end{table}

\FloatBarrier

\section{Determining the resistivity of each layer}
To find the resistivity of the individual layers, the resistivity of a Fe (2 nm)/(Ga,Mn)As (10 nm) patterned Hall bar was measured before and after the Fe and Al capping layer were chemically removed. To remove the oxidised Al capping layer, the sample was immersed for 15 s in a solution of 1:10 HCl:H$_2$O. The Fe layer was then subsequently dissolved by immersion for 15 s in MF319 photodeveloper. The change in sheet resistance from 523 $\Omega/$sq to 4534 $\Omega/$sq after removal of the metal layers, where the (Ga,Mn)As layer is 10 nm thick, indicates that only 12\% of the microwave current flows through the (Ga,Mn)As layer. The equivalent proportion for the 20 nm layers is 21\%.


\end{document}